\documentclass[20pt,aps,graphicx,preprint]{revtex4-1}
\usepackage{graphicx}
\usepackage{amssymb}
\usepackage{subfigure}
\begin{document}

\title{Formation of double ring patterns on Co$_2$MnSi Heusler alloy thin film by anodic oxidation
under scanning probe microscope}
\author{Vijaykumar Toutam$^{1,a)}$, Himanshu Pandey$^2$, Sandeep Singh$^1$ and R. C. Budhani}
\affiliation{National Physical Laboratory, Council of Scientific
and Industrial Research, New Delhi - 110012, India.}
\affiliation{Condensed Matter - Low Dimensional Systems
Laboratory, Department of Physics, Indian Institute of Technology,
Kanpur - 208016, India}
\email{vijay$\_$toutam@yahoo.co.in}
\begin{abstract}
Double ring formation on Co$_2$MnSi (CMS) films is observed at
electrical breakdown voltage during local anodic oxidation using
atomic force microscope (AFM). Corona effect and segregation of
cobalt in the vicinity of the rings is studied using magnetic
force microscopy and energy dispersive spectroscopy. Double ring
formation is explained on the basis of the interaction of ablated
material with the induced magnetic field of the tip to sample
current. Steepness of forward bias transport characteristics from
the unperturbed region of the CMS film suggest a non equilibrium
spin contribution, compared to corona region of the ring.
Formation of such mesoscopic textures in magnetic films by AFM tip
can be potentially used for lithography and memory storage
applications.
\end{abstract}
\maketitle
Formation of nanoscale oxide patterns on metal and semiconductor
surfaces by oxidation using a conductive scanning probe under
atomic force microscope (AFM) is termed as Local anodic oxidation
(LAO). This phenomena was first observed by Dagata
\textit{et.al}\cite{1} on Si (111), followed by similar
experiments on several bulk metal surfaces\cite{2,3},
semiconductors \cite{4,5} and supported thin films.\cite{6,7,8}
All these experiments have suggested that the electrochemical
reaction between the AFM tip and the surface under water meniscus
is very crucial for the formation of oxide patterns.\cite{9,10}
This phenomena is also observed under reactive conditions with a
suitable organic meniscus instead of a water meniscus.\cite{11}
Corona effect along the patterns during LAO was reported by few
groups.\cite{12,13} Corona formation was believed to be the effect
of the lateral diffusion of the oxyanions (OH$^-$) under high
percentage of humidity. For higher bias, a different mechanism
where a transient shock wave assisted ion spreading of OH$^-$ is
proposed.\cite{14} Intermixing of elements in a bilayer system of
GaAs/AlAs has also been observed during LAO.\cite{15} This is
supposed to be due to the electric breakdown of thin
semiconducting and dielectric films under anodic voltages greater
than 10 V.\cite{16} This process is very stochastic and is
explained by different mechanisms and simulations. Dielectric
breakdown is commonly studied for metal-insulator-metal
configurations to measure dielectric strength. Such studies have
generated considerable interest for resistive random access memory
devices. Many groups have demonstrated the reproducibility of
forming reversible metallic filaments at soft dielectric breakdown
leading to conductive paths which have great potential for high
density storage.\cite{17,18} Recently the possibility of inducing
local magnetic anisotropy using  relativistic energy and high
frequency e-beam has generated considerable interest for spin
electronics and memory applications.\cite{19}

Half metallic-ferromagnet full-Heusler alloy Co$_2$MnSi (CMS) is a
promising spintronic material due to its high Curie temperature
(~985 K)\cite{20,21} and theoretically predicted  100\% spin
polarization of conduction electrons.\cite{22} But experimentally
measured degree of polarization is $\approx$ 60\%.\cite{23} For
thin films of these materials, half metallicity is very sensitive
to the nature of surface and interface. It is demonstrated that
Mn-Mn termination retains the half-metallicity where as Co or
Mn-Si termination leads to mixing of spin sub bands.\cite{24}

In this paper, we report our results on AFM tip based anodic
oxidation of CMS and the effect of induced magnetic field on
pattern formation during dielectric breakdown. We explain how the
Oersted field generated is quite enough in magnitude to rearrange
the material, changing the properties of the heusler alloy
locally. Till now the Oersteds field generated during LAO has not
been given much importance as most of the studies were done on non
magnetic semiconductors and metal surfaces. This work was
initiated with the objective of creating planar nanostructures of
CMS films under AFM to study planar tunnel junctions and magneto
transport in nanowires of this half metallic compound. The process
of anodic oxidation in AFM, however, leads to the formation of
interesting ring structures. This paper describes our studies of
the topography, chemical composition and mechanism of formation of
such rings. We explain how the observed ring formation is
different from the general corona effect seen during LAO at high
voltages and explain the same by monitoring Conductive Atomic
Force Microscopy (CAFM) characteristics from different regions of
the affected area. Transport behavior of the formed junctions
under CAFM has same elements of spin diode characteristics. These
results are explained in analogy with spin polarized electrical
transport.

Co$_2$MnSi thin films were grown on (110) oriented SrTiO$_3$
substrates with Pulsed Laser Deposition (PLD) technique using KrF
excimer laser ($\lambda$ $\approx$ 248 nm).\cite{25} A typical
growth rate of ~ 0.56 {\AA}/s was used to deposit 40 nm thick
films at 200$^{o}$C, which were subsequently annealed at 400, 500
and 600$^{o}$C for one hour. The crystallographic structure of the
films was characterized using a PANalytical X'Pert PRO X-ray
diffractometer equipped with a CuK$_{\alpha1}$ source. Vibrating
Sample Magnetometer (VSM) with a maximum field of 1.7 Tesla was
used for in-plane magnetization measurements at room temperature.
More detail on structural and magnetic characterization are given
in supplemental material.\cite{Supple} A Multimode AFM with
Nanoscope V controller, (Veeco Ltd, USA) is used for all AFM
studies. All three samples were analyzed for their roughness. For
electrical breakdown and ring formation studies, a program written
in C++ interfaced with the controller is executed in Lithography
mode. The metal coated AFM tip was biased with respect to the
grounded sample and the experiment was carried out in ambient at a
relative humidity of 40\%. Lithography was done in contact mode
with -12 V of tip bias and tip velocity of 0.2 $\mu$m/s. For
Kelvin probe force microscopy, a metal coated tip biased at ac-
amplitude of 2.5 V and a frequency $\approx$ 80 kHz was used in
interleave scan, potential mode with a lift height of 50 nm. For
local I-V measurements and current mapping, an extended Tunneling
Atomic Force Microscopy (TUNA) module, Bruker AXS was used. For
conductivity mapping the current sensitivity of the amplifier was
set to 10 pA/V and scanned at 3V. Whereas for I-V, point
spectroscopy, current sensitivity factor is set to 1nA/V. Scanning
electron microscopy (SEM) imaging of the modified CMS region was
done using Ziess EO MA10 variable pressure SEM in conjunction with
energy dispersive spectroscopy (EDS) facility (Oxford Inca energy
250, Oxford instruments). CMS-Co junction was formed under AFM by
engaging Co coated AFM tip onto CMS film. For magnetic imaging,
Co/Cr coated AFM tip is magnetized and used in interleave scan
with a lift height of 30 nm.

Figure 1(a-c) shows AFM topographic images of all three films, CMS
400, CMS 500 and CMS 600 respectively, before and after the
application of bias. From these micrographs we conclude that while
surfaces of 400 and 500$^{0}$C annealed films are smooth,
600$^{0}$C annealing leads to roughening due to extensive crystal
growth. The average roughness of these films is 0.1 nm, 0.2 nm and
0.5 nm respectively. Figure 1(d-f) show the ring formation under
applied bias between AFM tip and CMS films annealed at 400, 500
and 600$^{0}$C respectively.
\begin{figure}
\begin{center}
\includegraphics [width=8 cm]{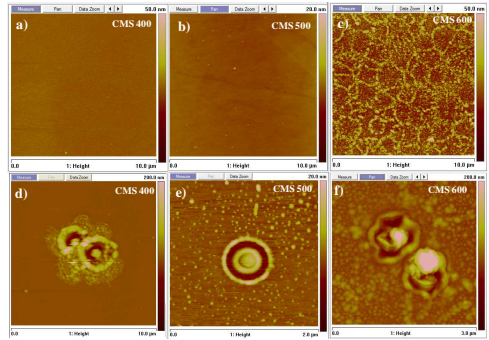}
\end{center}
\caption {\label{FIG 1} (a-c) Topography scan of CMS 400, 500 and
600 surfaces. (d-f) Double ring formation during local anodic
oxidation at -12 V on CMS 400, 500 and 600 surfaces respectively.
Spread of the material around the ring formation, defined as
Corona effect is observed for CMS 400 film.}
\end{figure}

CMS 400 forms the double ring with some asymmetry and the outer
diameter is of the order of 5 $\mu$m. Along with the ring
formation, a spread of the material in the vicinity of the
modified region which is identified as a corona effect is seen on
these films. Whereas CMS 600 forms highly irregular shaped
features, often it was found difficult to modify the CMS 600
surface. CMS 500 surface consistently gives a highly symmetric
double ring structure with least spread to the surroundings. No
corona effect is observed for CMS 500 and 600 films. Figure 2
shows a cross sectional profile across the diameter of one such
ring produced on CMS 500 surface. The double ring formation is
highly symmetrical as seen in Fig 2(a). The diameter of the outer
ring as measured from line profile shown in Fig. 2(b), is 1.1
$\mu$m and 0.6 $\mu$m for the inner ring. The height of the rim of
the ring is $\approx$ 12 nm. The AFM image and the shaded region
in the line profile prove that the material from the region
between the two rings is being ablated during the application of
bias. Even surface in the vicinity of ring is modified. This
represents the phenomena of stress induced by the biased AFM tip
on the surface and electrical break down of the film, under
intense electric fields. To evaluate whether the material has
undergone electrical breakdown, we have done the KPFM analysis of
one such modified region of CMS 400 film as shown in figure 4b
along with the topography of the film in Fig. 3(a). In KPFM, the
contact potential difference (CPD) betweem the metal tip and
sample surface is estimated based on Fermi level equilibrium
model.\cite{27} This is the amount of voltage required to nullify
the electrostatic force generated between two materials due to
their work function difference. The lower CPD in the modified
region as seen in Fig. 3(b) is a clear indication that negative
charge is induced in the modified region and its origin is the
dielectric breakdown of CMS films.\cite{28}

\begin{figure}
\begin{center}
\includegraphics [width=8 cm]{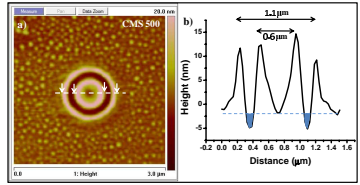}
\end{center}
\caption {\label{FIG 2} Double ring formation on CMS 500 film; a)
Topographic image, dielectric breakdown induced stress on the film
can be seen in the background. No corona formation is observed. b)
Line profile showing the cross section of the ring (shaded region
depicts the ablated material from the surface).}
\end{figure}

The pattern formed is a clear signature of dielectric breakdown.
As crystallinity of films improve with T$_A$, the threshold for
electrical breakdown also increases due to reduction in number of
traps.\cite{29} This agrees well with the trend observed in Fig. 1
in comparison with X-ray diffraction data shown in Fig. S1. Hence
CMS 600 film of high crystalline quality is unlikely to undergo
easy electrical breakdown where as CMS 400 can be ablated easily
and the material spread to the surroundings. However, the
formation of symmetric ring features on CMS films is very distinct
and intriguing. To understand any contribution from the magnetic
nature of these films and interaction of the ablated material with
the magnetic field generated by the conducting tip, the ring and
their surroundings formed on CMS 400 thin film were mapped for
their magnetic nature by magnetic force microscopy (MFM). Energy
dispersive spectroscopy (EDS) was also done for elemental analysis
of various regions of the ring. Both these results unequivocally
showed that the distributed material is magnetic in nature and
mostly rich in Co. Figure 4 summarizes the results of MFM and EDS
analysis of the ring and its corona. The AFM topographic image and
the MFM image at a lift height of 30 nm, collected from the same
area are displayed in Fig. 4(a) and 4(b) respectively. The MFM
images have a phase variation different from the rest of the film
explaining the magnetic nature of the dispersed material. Clearly,
the intermittent out of phase variation of outer rim and inner rim
of the double ring, pointed by arrow marks in MFM image i.e., Fig.
4(b) explains the non magnetic behavior. This negative phase
variation could be due to the local anodic oxide formation along
the rim of the CMS rings.
\begin{figure}
\begin{center}
\includegraphics [width=7 cm]{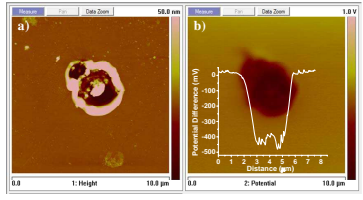}
\end{center}
\caption {\label{FIG 3} : KPFM of CMS double ring a) Topography of
the double ring formed under break down at -12 V of tip bias. b)
CPD image of the same region using KPFM mode. Lowering of contact
potential difference by 0.5 V in the break down region}
\end{figure}
Figure 4d shows the energy dispersive spectrum of different
elements from the CMS film. The inset shows the variation in
atomic \% of Co for different regions of the SEM image in Fig.
4(c). Region 1 corresponds to the ablated portion of the ring,
region 2 is the corona portion of the ring and region 3 is the
unmodified substrate region. High atomic \% of Co in the vicinity
of the ablated region compared to rest of the film explains the
enhanced magnetic contrast of MFM image in figure 5b. This clearly
says that the pattern formation and the spread of the material on
CMS films are due to the interaction of Oersted's field with the
material itself. This is a distinct phenomenon compared to general
corona effect observed during LAO of semiconductors and metals.
The electric field intensity (E) at the edge of the AFM tip for a
bias of -12 volts is of the order$\approx$ $10^7$ V/cm.\cite{30}
The corresponding current density (J) can be estimated by field
emission phenomena based on Fowler-Nordhiem theory (F-N
theory).\cite{31}

\begin{figure}
\begin{center}
\includegraphics [width=7 cm]{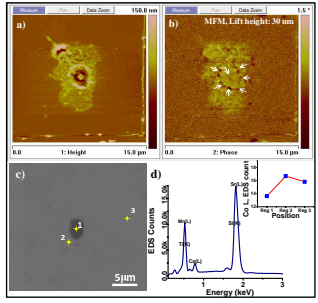}
\end{center}
\caption {\label{FIG 4} Magnetic nature of the distributed
material around the ring on CMS 400 film probed using Magnetic
force microscopy and Energy dispersive spectroscopy. a) Topography
of the ring formation with corona; b) MFM image of the same at a
lift height of 30 nm; c) SEM image of the CMS ring; d) Plot
showing variation of Co atomic $\%$ in three different regions
shown in Fig 5c.}
\end{figure}

\begin{figure}
\begin{center}
\includegraphics [width=7 cm]{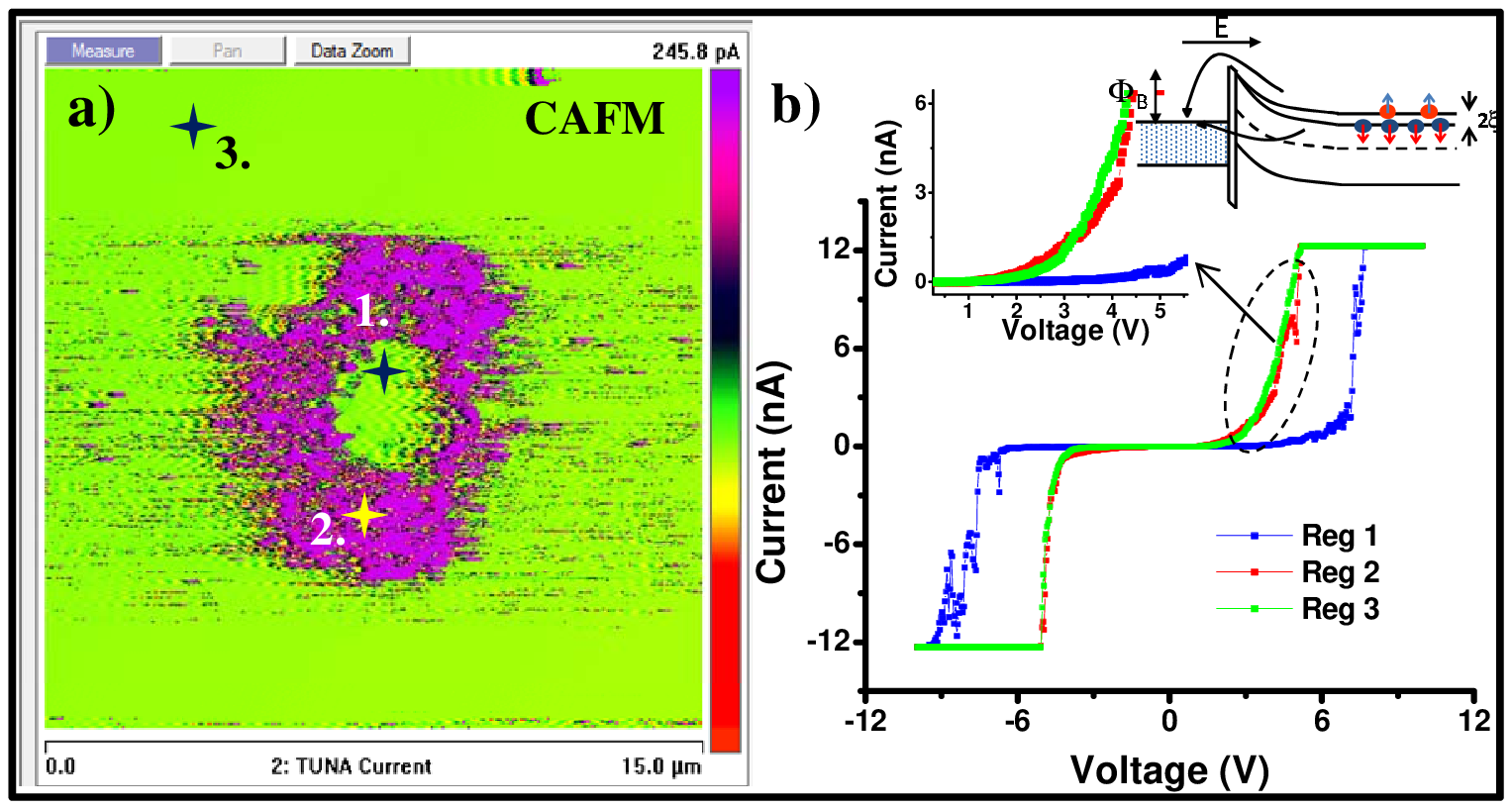}
\end{center}
\caption {\label{FIG 5} (a) Current mapping of the ring formed on
CMS 400 sample; (b) I-V captured in the rim and on the periphery
measuring the variation in conductivity. The forward bias I-V
characteristics for bare film is more sensitive compared to the
corona region.}
\end{figure}
Figure 5 shows the conductivity mapping of the double ring on CMS
400 film at tip voltage of $\approx$ 3 V, along with
current-voltage characteristics from different regions captured by
point and shoot I-V spectroscopy mode. The contrast of the current
map in Fig. 5(a) reveals that the conductivity of corona region
(Region 2) is more compared to that of the bare region (Region 3)
and the rim region (Region 1). Figure 5(b) shows the I-V
characteristics captured from these three different regions of the
ring structure along with an inset showing the section of I-V
characteristics and the schematic of electron transport. From the
I-V curves it is clear that the conductance of the rim region
(Region 1) is less than that of the Region 2 and 3 which are taken
from the corona and bare CMS respectively, proving that it is
mostly oxide. The behavior of I-V characteristics are typical of a
schottky junction. For such junctions, current in forward bias is
dominated by majority charge carriers. Modified I-V
characteristics for schottky junction in the present case can be
expressed as;
\begin{eqnarray}
I = I_S \left\{ {e^{\frac{{ev}}{{kT}}} \left( {1 + \delta P} \right) - 1} \right\},I_S  = AA^* T^2
e^{ - \frac{{\phi _B }}{{kT}}}
\end{eqnarray}
where \textit{I$_S$} is the saturation current, \textit{B} is
Schottky barrier height, $\delta$P is the non equilibrium spin
polarization in the n region, v is applied voltage, A is area, and
A$^\star$  is Richardson's constant. The current map in Fig. 5(a)
agrees well with the I-V characteristics below 3 V from Region 2
and Region 3 as shown in the inset of Fig. 5(b).  Difference in
the steepness of I-V characteristics in the forward bias from both
regions above 3 V gives an indication that the current from Region
3 has a contribution not only from charge drift but also from
spin, making it behave like a Spin diode.\cite{34} The transport
characteristics in forward bias is similar to the current
injection in spin diode, where the I-V characteristics are
expressed as\cite{35}:
\begin{eqnarray}
I = I_0 \left\{ {e^{\frac{{ev}}{{kT}}} \left( {1 + \delta P_n
\cdot P_p } \right) - 1} \right\}
\end{eqnarray}
where P$_p$ is equilibrium spin polarization in the p region. When
unpolarized current is passed across a ferromagnetic
semiconductor, the current becomes spin-polarized.\cite{36,37}
Electrons which are spin polarized in this material could show a
strong interaction with the applied current. Transverse spin
current has been recently demonstrated experimentally in GaAs thin
films,\cite{38} and in AlGaAs quantum wells.\cite{39} The steeper
I-V characteristics for bare region compared to modified region
indicates the spin contribution in the transport and loss of half
metallicity of the modified region due to electrical breakdown.

Double ring structures were formed on Co$_2$MnSi films via
electric breakdown, using conducting AFM tip. These rings were
analyzed using magnetic force microscopy and energy dispersive
spectroscopy. MFM of the rings suggest that the dispersed material
forming a corona is magnetic and EDS confirmed that the material
is rich in Co. The ring formation is explained in terms of the
interaction of ablated material with the magnetic field produced
by the conducting AFM tip. Local I-V spectroscopy using Co tip on
the ring and its surrounding regions suggest that the system forms
a schottky diode when in contact with the tip. The I-V
characterization from bare region indicate that the transport is
influenced by the spin polarization nature of conduction electrons
in CMS. Further experiments of I-V characterization at different
temperatures and in the presence of magnetic field will confirm
the observation of spin transport of the system.

VKT thanks Dr. Sukhwir Singh for his support and cooperation in
accessing the AFM facility, Mr. K. N. Sood for getting SEM data
and Dr. S. T. Lakshmi Kumar for his valuable comments about the
work. H.P. acknowledges financial support from Indian Institute of
Technology (IIT) Kanpur and the Council for Scientific and
Industrial Research (CSIR), Government of India.  R.C.B.
acknowledges the J. C. Bose fellowship of Department of Science
and Technology. The Work carried out at IIT Kanpur has been
supported by the Department of Information Technology.

\end{document}